\documentstyle[prl,aps]{revtex}
\begin{document}
\twocolumn[\hsize\textwidth\columnwidth\hsize
\csname@twocolumnfalse\endcsname

\draft      
\title{Confinement of acoustical vibrations in a semiconductor planar phonon cavity}
\author{M. Trigo$^1$, A. Bruchhausen$^1$, A. Fainstein$^1$,  B. Jusserand$^2$, 
and V. Thierry-Mieg$^2$}     
\address{$^1$Centro At\'omico Bariloche \& Instituto Balseiro, C.N.E.A., 
8400 S. C. de Bariloche, R. N., Argentina}       
\address{$^2$Laboratoire de Photonique et de Nanostructures, CNRS, Route de Nozay, 
91460 Marcoussis, France}
                                                                    
\maketitle                                                                      
\begin{abstract}                                                                
Extending the idea of 
optical microcavities to sound waves, we propose a {\it phonon cavity} consisting
of two semiconductor superlattices enclosing a spacer with thickness determined by the 
acoustic wavelength at the center of the first zone-center folded 
minigap. We show that acoustical phonons can be confined in these layered structures,
and propose Raman experiments which are able to probe these novel 
excitations. The Raman experiments take profit of an optical microcavity scattering geometry 
that, through the forward-scattering contribution, gives access to the zone-center 
excitations. We report experimental results of Raman scattering in a structure
based in GaAs/AlAs materials that demonstrate unambiguously the observation of
phonon cavity confined acoustical vibrations. The experimental results compare 
precisely with photoelastic model calculations of the Raman spectra of the 
proposed phonon-cavity embedded optical microcavity.
\end{abstract}    

\vskip 2pc]\narrowtext                                                                    

The wave character of electrons and their interaction with a periodic 
potential in crystals leads to the Bragg reflection and opening of
forbidden energy gaps.~\cite{Aschcroft} An impurity state with energy
within this gap is reflected back in all three propagation directions
and thus its wavefunction is confined in space. The equivalent of a
periodic potential for optical waves (i.e., photons) is a periodic 
structure of materials with contrasting refractive index. A planar 
periodic stack of two materials $\lambda/4$ thick reflects photons
propagating normal to the layers within a stop-band around the wavelength 
$\lambda$, and is termed a ``distributed Bragg reflector" (DBR).~\cite{Yeh}
Photon gaps can also be opened in the three space directions by 
appropriately selecting the dielectric material and structure.\cite{Yablonovitch}
A planar microcavity is a spacer of thickness $\lambda/2$  enclosed
by two DBR's.~\cite{Weisbuch} Photons of wavelength $\lambda$ are confined in such
structure much like an electronic defect state in a crystal.\cite{Stanley}
This photon confinement leads to fundamental changes in the light-matter
interaction.\cite{Weisbuch}
Phonons are vibrational waves described by similar wave equations as 
photons, but which are subject to mechanical (instead of electromagnetic)
boundary conditions at the interfaces.
Extending the above ideas to vibrational waves, in this letter we propose a 
planar ``phonon cavity" structure designed to confine acoustical phonons. 
In addition, we show theoretically that Raman scattering through an optical
cavity geometry~\cite{Fainstein1,Fainstein2} is able to probe these 
novel excitations, and we demonstrate their existence through
experiments in a real GaAs/AlAs based structure. These results
are relevant to diverse phenomena as, e.g.,  phonon amplification and 
stimulated emission\cite{Tucker,Haken,Fokker}, 
coherent generation and control of phonons\cite{Dekorsy}, and
modified electron-phonon interactions.

The acoustical phonon branches in a superlattice (SL) made of
a periodic sequence of semiconductor layers can be described by 
backfolding the phonon dispersion of an average bulk solid and
opening of small minigaps at the zone-center and reduced
new Brillouin-zone edge.~\cite{Jusserand1}
It has been proposed long time ago that the minigap of these zone-folded
SL acoustical phonons can be used as a ``phonon filter'', 
acting very much like dielectric DBR's but for the selective reflection
of high frequency sound waves.~\cite{Narayanamurti}
Thus, a planar ``phonon cavity" can be constructed by enclosing
between two SL's a spacer of thickness $\lambda_{ac}/2$, where
now $\lambda_{ac}$ is the wavelength of the acoustical phonon at
the center of the phonon minigap. In Fig.\ref{fig1} we present a
scheme of the proposed phonon cavity structure, based in semiconductor 
GaAs and AlAs materials. The ``phonon DBR's" consist of 11 period
$74\AA/38\AA$ GaAs/AlAs SL's, and enclose a $50\AA$ Al$_{0.8}$Ga$_{0.2}$As
spacer.

The acoustical phonons in the proposed layered structure can be evaluated
using a matrix method implementation of the standard elastic 
continuum description of sound waves originally proposed by Rytov.~\cite{Rytov} 
With this model one can obtain the 
transmission of acoustical waves through the structure, and the
displacement pattern as a function of the phonon energy. 
In Fig.\ref{fig2}(top) we display the calculated phonon reflectivity
for energies around the first zone-center minigap. 
For this calculation standard material parameters (sound velocities and 
densities) were used.~\cite{Adachi}
A phonon-transmission
stop-band is observed, coincident with the SL's minigap. Within 
this stop-band a phonon-cavity mode exists, characterized by complete
transmission of the vibrational energy. The calculated phonon amplitude 
for this latter mode is shown in Fig.\ref{fig3}. While phonon modes with
energies within
the SL's minigap decay exponentially in space, the phonon-cavity
mode propagates through the structure, and its intensity is enhanced within the
cavity spacer.

The design of the phonon-cavity deserves some
consideration. In principle, any of the acoustical minigaps
can be used as phonon DBR. The center of the phonon stop-band (and
hence the energy of the confined mode) is simply determined
by the SL's period $d=d_1+d_2$. The stop-band width, on the other hand, 
is proportional to a modulation parameter $\epsilon$ and displays 
an oscillatory behavior as a function of the SL layers thickness 
ratio $d_1/d_2$.\cite{Jusserand1} 
Here $\epsilon=(\eta_2-\eta_1)/(\eta_1\eta_2)^{1/2}$,
with $\eta_i=\rho_i v_i$, $\rho_i$ and $v_i$ the acoustic impedance,
density and sound speed of layer $i$, respectively.  
The oscillatory function, and thus the thickness ratio 
$d_1/d_2$ appropriate for the DBR's
of a phonon-cavity, depends also on the order of the minigap $\nu$.
For the first zone-edge minigap ($\nu=1$) the cavity is optimized
using $d_1/v_1=d_2/v_2=d_c/2v_c$, while for the first
zone-center minigap ($\nu=2$) the relation is  
$d_1/v_1=d_2/3v_2=d_c/2v_c$. Here $d_c(v_c)$ is the 
cavity thickness (sound velocity). These expressions can
be directly transferred to optical waves by replacing
the acoustic impedances in $\epsilon$ by the refractive
indices $n_i$. In fact, for GaAs and AlAs it turns out
that $\epsilon$ is basically the 
same for phonons and photons (0.18 and 0.175, respectively),
implying that similar DBR reflectivities and cavity
$Q$-factors can be obtained as a function of the number
of DBR layers ($N$). For our phonon cavity, designed to operate at
the first zone-center minigap with 11 SL periods,
$Q\sim 140$. On the other hand, $Q$-factors in the range 
$Q > 3000$ can be easily obtained with $N\sim20$.

The issue we address next is how to create
and study the confined vibrational excitations in phonon cavities. 
Superconducting tunnel junctions
have been used in the past as sources and detectors of high-frequency 
phonons with, however, limited resolution.~\cite{Narayanamurti} As derives
from Fig.~\ref{fig2}, we require a technique able to resolve $\sim 0.1$~cm$^{-1}$
(i.e., $~10^{-2}$~meV) in the energy range $\sim 5-30$~cm$^{-1}$ (1-4 meV). 
We  propose Raman experiments based in a planar microcavity scattering geometry 
that, besides providing the required resolution~\cite{Jusserand1} 
and an amplified efficiency ($\sim \times 10^{5}$),~\cite{Fainstein1,Fainstein2}
give access to the zone-center excitations through the forward-scattering 
contribution.\cite{Fainstein2} As we will
show  next, these singular characteristics  allow
the observation of these novel acoustical phonon
excitations.

We consider a phonon-cavity embedded optical microcavities as depicted
in Fig.\ref{fig1}. We have performed calculations of the Raman spectra of the proposed
structure based on a photoelastic model for
the Raman efficiency. Within a photoelastic model for scattering by longitudinal acoustic 
phonons the Raman efficiency is given by
$I(\omega_{q_z}) \propto \left| \int dz E_L E^*_S p(z) 
\frac{\partial \Phi(z)}{\partial z}
\right|^2$,
where $E_L$($E_S$) is the laser(scattered) field, $p(z)$ is the spatially
varying photoelastic constant, and $\Phi$ describes the (normalized) phonon
displacement.~\cite{Jusserand1} For an infinite periodic SL the wavevector $q_z$ is a good
quantum number. For a standard Raman scattering geometry, 
with laser and scattered fields given by plane waves ($E_L E^*_S=e^{(k_L-k_S)z}$),
$I(\omega_{q_z})$ leads to the usual phonon doublets
with transferred wavevector given by the conservation law $q_z=k_L-k_S$.
For the case we are discussing, however, $q_z$ is only partially conserved
due to the finite-size of the structure.~\cite{Fainstein1,Trigo} In addition,
$E_L$ and $E_S$ correspond to the cavity confined photons.
Under double 
optical resonance conditions (that is, both laser and Stokes fields resonant with
cavity modes),\cite{Fainstein1} the incident and scattered photon fields 
are given within the optical cavity spacer by the same standing wave 
$E_L=E_S=e^{ik_zz}+e^{-ik_zz}$ 
(here, by construction, $k_z=\pi/D$ with $D$ the spacer width).~\cite{Fainstein2}
Consequently, $E_L E^*_S \propto \left|E(z)\right|^2=2+
\left(e^{i2k_zz}+e^{-i2k_zz}\right)$.
Thus, both a forward scattering (the term 2) and a backscattering contribution
($e^{i2k_zz}+e^{-i2k_zz}$) are coherently added to the Raman
efficiency $I(\omega_{q_z})$. The former corresponds to a wavevector transfer
$q_z=0$, while for the latter $q_z=2\pi/D$.

We show in Fig.\ref{fig2}, together with the phonon
reflectivity curve, a calculated Raman spectrum for the phonon-cavity
embedded photon-cavity of Fig.\ref{fig1}. For this calculation
the laser wavelength is $\lambda=833$~nm, the layer´s refractive 
indices $n_{GaAs}=3.56$ and $n_{AlAs}=2.97$
and, being $\lambda$ in the transparency region, the photoelastic
constants are taken real.\cite{Jusserand1,Adachi}
For discussion purposes we also present, in the bottom
panel, the separate contributions of the forward scattering (FS) and
backscattering (BS) terms. Note that the main 
Raman peak in Fig.\ref{fig2} corresponds to scattering by the 
phonon-cavity confined  mode. Interestingly, this latter feature comes 
{\em only} from the FS ($q_z=0$) contribution. Its relatively
large intensity is explained by the spatial confinement of the
cavity-phonon mode at the same place where the confined cavity {\em photon}
has the largest amplitude. In addition, several other 
features should be noted in the calculated spectra:
$(i)$ besides the confined phonon mode, the
FS contribution displays scattering at
the low energy side of the minigap, corresponding to the
(-1) zone-center SL folded minigap mode.\cite{Jusserand1} 
The (+1) peak is forbidden in an infinite SL due to a parity 
selection rule,\cite{Jusserand1} and is only weakly perceptible
in the shown spectra due to a partial relaxation of
$q_z$-conservation due to the phonon-cavity SL mirror's
finite-size. $(ii)$ The BS contribution, on the other hand, 
displays the usual SL phonon doublet. $(iii)$ Side oscillations 
are observed that modulate the spectra. These oscillations
originate in the SL's finite-size,~\cite{Giehler,Lockwood} and are
particularly intense in the cavity spectrum because they add 
coherently from the BS and FS terms. 
With ``coherence" here we mean that
the BS and FS contributions are added {\em before} (and not after)
squaring in $I(\omega_{q_z})$. And last, 
$(iv)$ the coherent addition of BS and FS in the
total cross section is particularly clear around $\sim 14$~cm$^{-1}$
where the cavity spectrum is not the 
simple sum of the individual contributions.

With the above Raman calculations in mind we fabricated, using
standard molecular beam epitaxy techniques, a phonon-cavity
embedded optical cavity with the structure and nominal
layer thicknesses as given in Fig.\ref{fig1}. 
The structure was purposely grown with
a slight taper which enables a tuning of the optical cavity mode
by displacing the laser spot on the sample surface. 
The layer widths were chosen so that the optical cavity mode 
falls below the GaAs gap, in order to avoid parasitic
luminescence coming from the substrate. 
The coupling of both laser and Stokes (non-degenerate) photons
with the optical cavity mode is accomplished by tuning
the incidence angle as described in Ref.~\onlinecite{Fainstein1}.
Since the acoustical folded-phonon-like excitations are in the range 
10-30 cm$^{-1}$ ({\em i.e.}, $\sim 1-4$ meV), incidence angles 
around $\sim 3-5^\circ$ are required. 
The Raman experiments were performed at 77~K using a triple
Jobin-Yvon T64000 spectrometer equipped with a N$_2$-cooled
charge-coupled-device. A Ti-sapphire laser was used as
the excitation source (power below 20 mW) at energies in 
the range 830-850~nm,
well below the SL's fundamental exciton absorption 
($\sim 1.6$~eV=775~nm) and GaAs gap (1.515~eV=820~nm).
The spectra were acquired using a triple subtractive 
configuration with spectral resolution around 
0.1-0.2~cm$^{-1}$.

We show in Fig.~\ref{fig4} a typical experimental spectrum
in the spectral range corresponding to the acoustical folded-phonon
first zone-center minigap and obtained with laser wavelength
$\lambda=833$~nm. The {\em optical} cavity mode linewidth, determined mainly by
the collection solid angle ($f/2.5$ optics) and the laser spot diameter, 
was around 10-12~cm$^{-1}$. This implies that no
peak is selectively amplified, and thus that the relative intensity of
the different spectral features is intrinsic.\cite{Fainstein1} 
To display only the Raman contribution a weak luminescence lorentzian 
background corresponding to the optical cavity mode was subtracted in 
the shown spectrum. Identical spectra were acquired from different 
spot positions and with varying laser energies. A main peak centered 
at $\sim 15$~cm$^{-1}$ is observed, together with secondary lines and
oscillations. In Fig.\ref{fig4}
we also show for comparison the calculated Raman curve corresponding to the
nominal structure. The agreement with the experimental curve
is remarkable {\em without any adjustable parameter}. This
includes the observation of the sought phonon-cavity mode,
but also almost every other detail of the spectra. In fact, the 
only apparent difference is the predicted (and non observed) 
splitting of the peak in the low energy side of the phonon minigap
($\sim 14.3$~cm$^{-1}$).
This latter peak corresponds to the (-1) component of the
zone-center doublet, mainly determined by the FS contribution
(see Fig.\ref{fig2}). Its
detailed shape is strongly dependent on the interference
between the FS and BS terms. As we will show next, the non-observation
of the peak splitting can be understood as due to a small but expectable 
disorder due to interface roughness.

We have considered three possible mechanisms to account for the non-observation
of the above discussed FS peak splitting, namely $(i)$ differences of the real structure
phonon-cavity spacer width with respect to the nominal value, $(ii)$ 
destructive interferences in the Raman cross section originated in 
differences of the widths of the two (otherwise perfect) SL's~\cite{Giehler} 
and, finally $(iii)$ interface roughness leading to width variations 
around the nominal value for all the layers making the two SL's. We 
found that only the last choice can explain the observed spectra
with reasonable values of disorder. This is shown 
in Fig.\ref{fig4}, with a curve obtained by adding 300 spectra corresponding
to phonon-cavities with SL layer widths with random disorder $\pm 4\%$
around the nominal values. This corresponds to changes of $\pm 2.8 \AA$, or
$\pm 1$ GaAs or AlAs atomic monolayers at the interfaces.  
All peaks besides the 14.3~cm$^{-1}$ line are quite stable 
to this disorder, and thus do not change except for a line
broadening which reproduces almost exactly the measured spectral
widths. 
The splitting of the 14.3~cm$^{-1}$ peak, on the other hand,
is washed out in correspondence with the experiment. 
In fact, as it is clear from Fig.\ref{fig4} the agreement between
theory and experimental results is striking when the interface
roughness is taken into account. This agreement includes the observation 
of all the main lines and most of the weaker oscillations, their 
spectral positions and peak widths.

In conclusion, we have proposed a phonon-cavity that displays novel
confined acoustical excitations. In addition, we showed through
photoelastic model calculations that these vibrations can be generated
and probed by Raman scattering experiments that exploit the enhancement and
confinement of photons in optical microcavities. Last, we fabricated
a phonon-cavity embedded optical microcavity based in GaAs and AlAs
materials, and succeeded in observing the cavity-confined high frequency
hyper-sound mode. Calculations that take into account a small but expectable
interface roughness are able to account precisely for every detail of the observed
spectra. The confined phonon modes in a phonon cavity are spatially localized
highly mono-energetic excitations. These features open the way to studies of 
phonon stimulation, coherent phonon generation, and modified (enhanced or inhibited) 
electron-phonon interactions in phonon cavities. The confinement and 
enhancement of photons in an optical microcavity can be exploited, in addition, 
for the optical generation of these confined phonon modes.

\newpage
\begin{figure}
\caption{
Scheme of the proposed phonon-cavity embedded optical cavity. 
The phonon ``mirrors"  consist of 11 period
$74\AA/38\AA$ GaAs/AlAs SL's, and enclose a $50\AA$ Al$_{0.8}$Ga$_{0.2}$As
spacer. The phonon-cavity structure, on the other hand, constitutes the $\lambda$ 
spacer of an optical microcavity enclosed by 20 DBR 
Ga$_{0.8}$Al$_{0.2}$As/AlAs pairs on the bottom and 16 on top.
\label{fig1}}
\end{figure}

\begin{figure}                                                                  
\caption{
Calculated phonon reflectivity (top) and Raman spectrum (bottom) 
in the spectral region around the first zone-center minigap for 
the phonon-cavity embedded photon-cavity depicted in Fig.\ref{fig1}.
The separate contributions of the forward scattering (FS) 
and backscattering (BS) terms is also shown.
\label{fig2}}                               
\end{figure}

\begin{figure}    
\caption{
Calculated phonon amplitude corresponding to the confined phonon-cavity mode.
The curve is shown superimposed to a scheme of the phonon structure
characterized by the material´s photoelastic constant.
\label{fig3}}                               
\end{figure}

\begin{figure}                                                                  
\caption{
Experimental Raman spectrum
in the spectral range corresponding to the acoustical folded-phonon
first center-zone minigap (Exp.), obtained with laser wavelength
$\lambda=833$~nm. Also shown are calculated Raman curves 
corresponding $(i)$ the nominal structure (Nominal), and $(ii)$ a 
phonon-cavity with random disorder $\pm 4\%$ of the
SL layer's thickness (Disorder). 
\label{fig4}}                               
\end{figure}

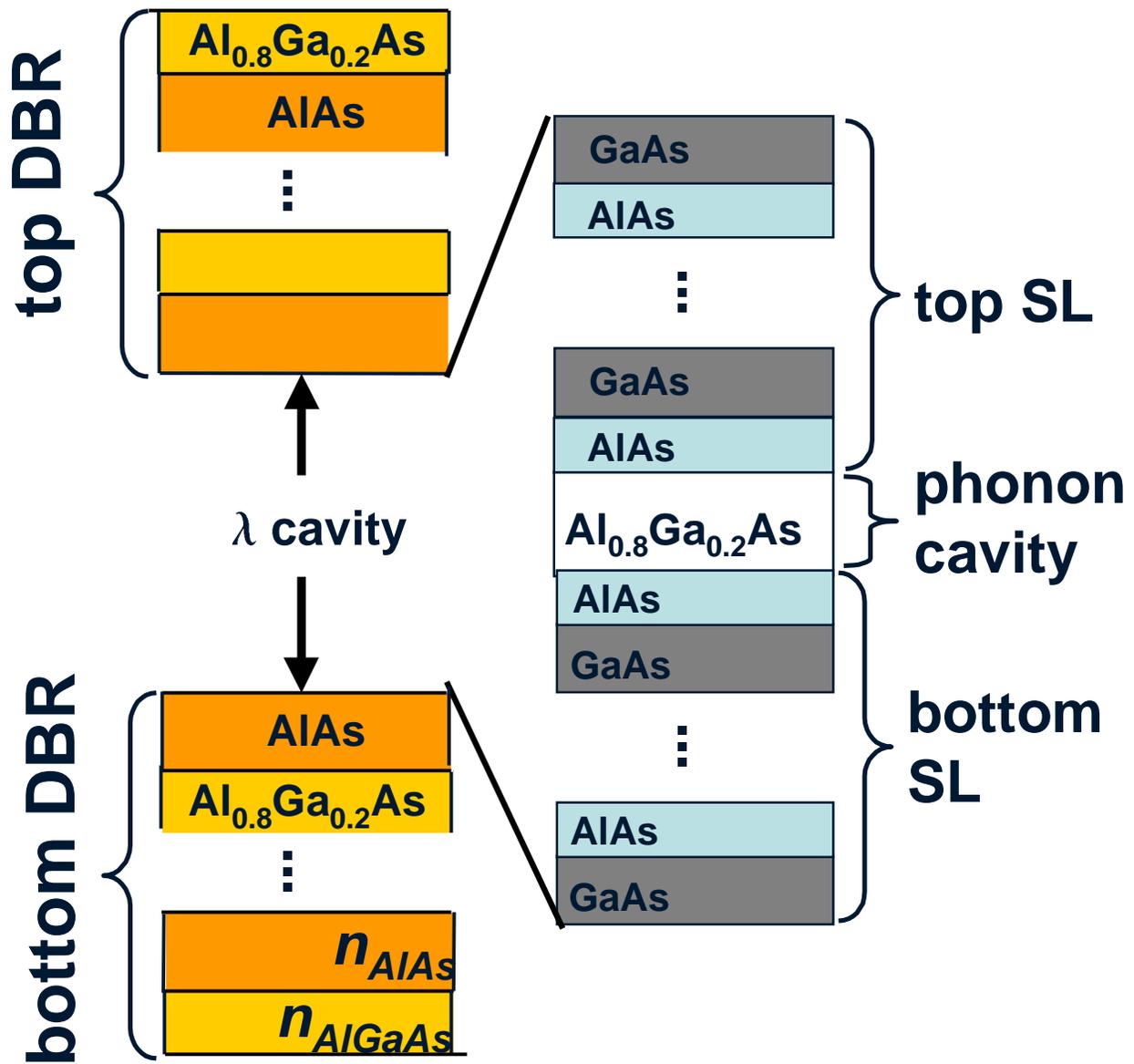
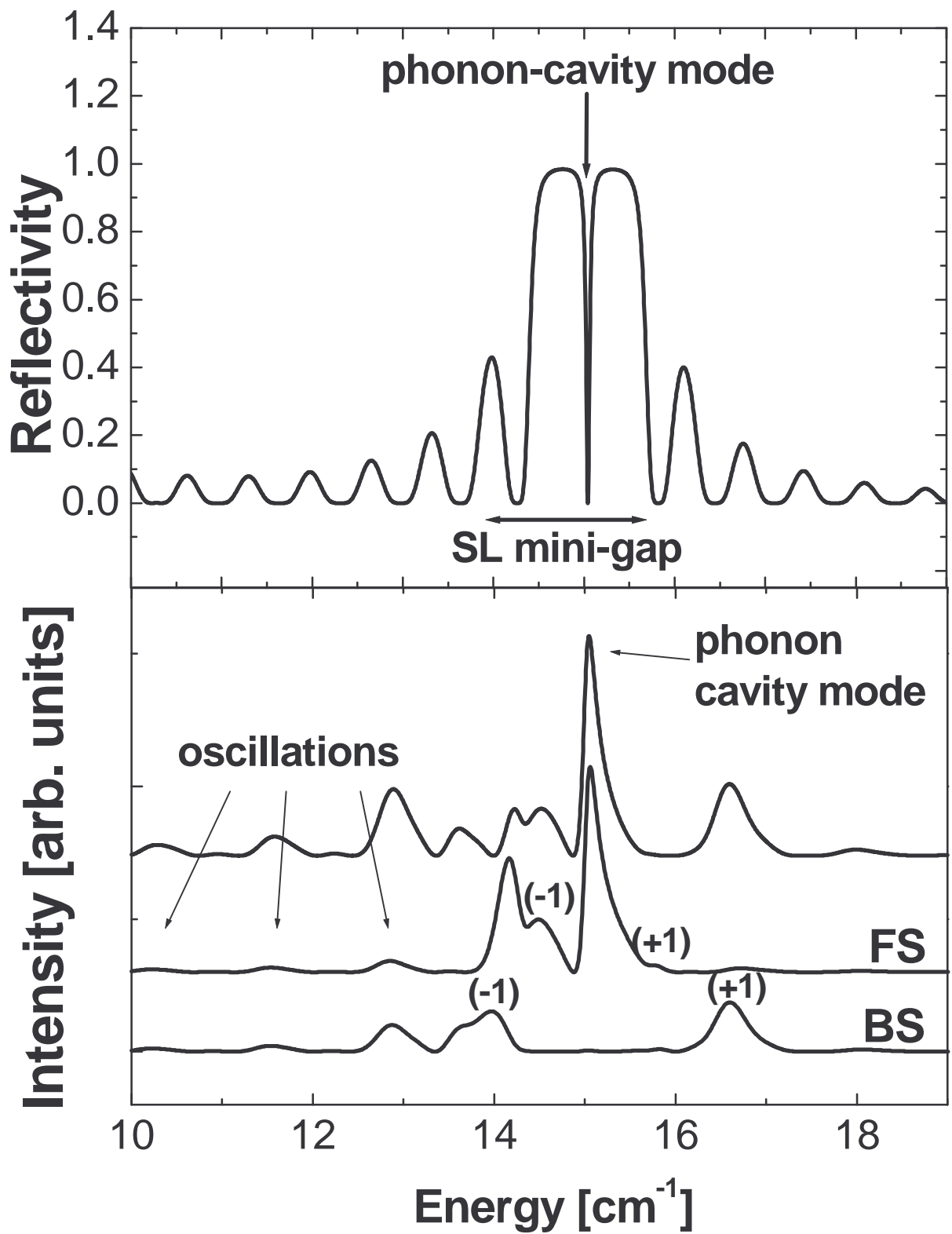
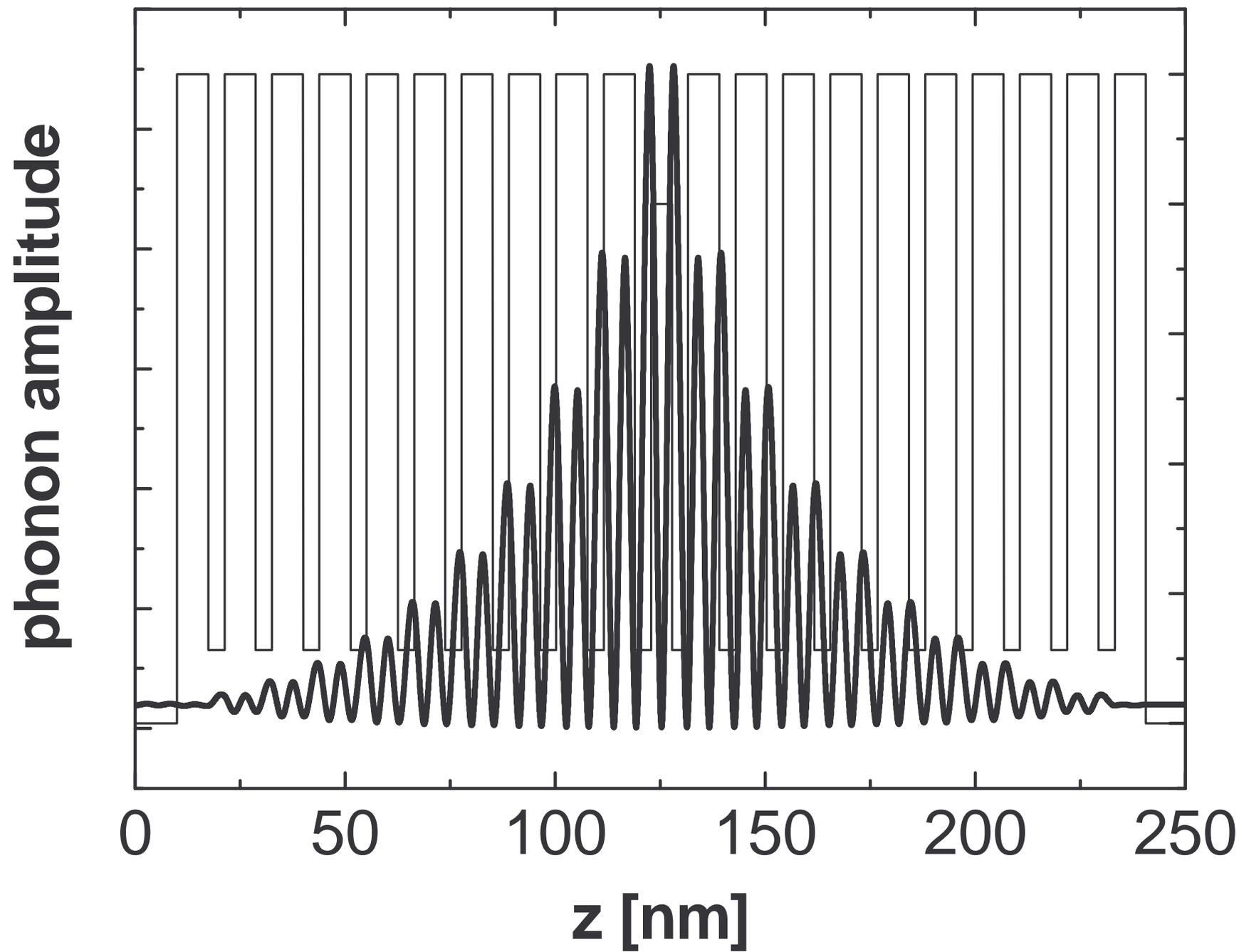
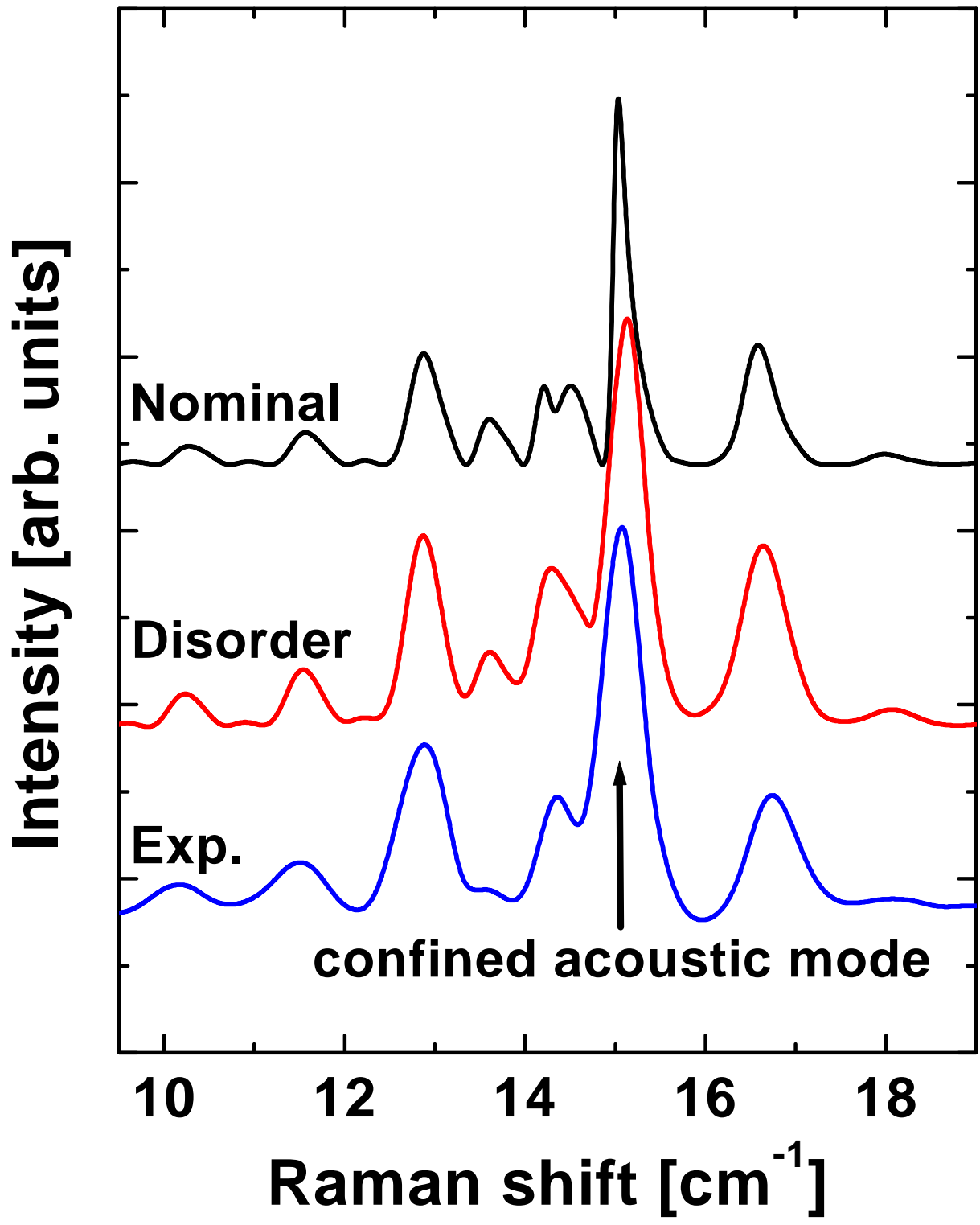
\end{document}